\documentclass[10pt,conference]{IEEEtran}

\usepackage{cite}
\usepackage{graphicx}
\usepackage{multirow}
\usepackage{makecell}

\usepackage{tikz}
\usepackage{amsmath}
\usepackage{amssymb}

\usepackage{float}
\usepackage[noend]{algorithm2e}

\usepackage{subcaption}

\AtBeginDocument{%
  \providecommand\BibTeX{{%
    \normalfont B\kern-0.5em{\scshape i\kern-0.1em b}\kern-0.1em\TeX}}}

\begin{document}
\title{A Neural Question Answering System for Basic Questions about Subroutines}

\author{\IEEEauthorblockN{Aakash Bansal\IEEEauthorrefmark{1},
		Zachary Eberhart\IEEEauthorrefmark{1}, Lingfei Wu\IEEEauthorrefmark{2} and
		Collin McMillan\IEEEauthorrefmark{1}\\
		\IEEEauthorblockA{\IEEEauthorrefmark{1}Dept. of Computer Science,
			University of Notre Dame, Notre Dame, IN, USA \\
			\{abansal1, zeberhar, cmc\}@nd.edu}
		\IEEEauthorblockA{\IEEEauthorrefmark{2}IBM Research,
			Yorktown Heights, NY, USA \\
			\{wuli\}@us.ibm.edu}}}

\maketitle

\begin{abstract}
A question answering (QA) system is a type of conversational AI that generates natural language answers to questions posed by human users.  QA systems often form the backbone of interactive dialogue systems, and have been studied extensively for a wide variety of tasks ranging from restaurant recommendations to medical diagnostics.  Dramatic progress has been made in recent years, especially from the use of encoder-decoder neural architectures trained with big data input.  In this paper, we take initial steps to bringing state-of-the-art neural QA technologies to Software Engineering applications by designing a context-based QA system for basic questions about subroutines.  We curate a training dataset of 10.9 million question/context/answer tuples based on rules we extract from recent empirical studies.  Then, we train a custom neural QA model with this dataset and evaluate the model in a study with professional programmers.  We demonstrate the strengths and weaknesses of the system, and lay the groundwork for its use in eventual dialogue systems for software engineering.
\end{abstract}



\vspace{-0.1cm}
\section{Introduction}

A question answering (QA) system is a type of conversational AI that focuses on generating natural language answers to questions posed by human users.  QA is defined as single-turn dialogue, in that there are only two participants in the conversation (the human and the machine) and each participant speaks for only one turn (the human asks a question which the machine answers).  In practice, a complete conversational machine agent would discuss several topics over an arbitrary number of turns, detect when a question has been asked, and the use a QA system to generate an answer to the question.  Thus, QA systems are key components necessary for building usable conversational agents.

In general, QA systems generate an answer given a context about which the question is being asked.  For example, Yin~\emph{et al.}~\cite{yin2016neural} describe an approach that parses a knowledge base of facts about famous people to generate English answers about birthdates, political offices held, awards received, etc.  Malinowski~\emph{et al.}~\cite{malinowski2015ask} present a system that answers questions about images, such as which objects are red or green in the image.  Weston~\emph{et al.}~\cite{weston2015towards} provide a dataset of twenty tasks for training QA systems (the so-called bAbI tasks) ranging from positional reasoning to path finding, for which the context is a knowledge base of facts about objects and how they relate to each other (e.g. Context: 1. Lily handed the baby to Philip. 2. Philip walked outside. Question: Where is the baby?  Answer: Outside with Philip.).

As the above examples show and as chronicled in several survey papers~\cite{liu2017survey, chen2017survey, gao2019neural}, scientific literature from the areas of Natural Language Processing (NLP) and AI is replete with QA systems designed to answer questions about a context.  The overall structure of these approaches is fairly consistent: A large dataset is collected including question, answers, and related contexts.  Then a model is trained and tested using the dataset.  Typically, an encoder-decoder neural network is employed, in which the model learns to connect features in the questions to features in the context via an attention mechanism.  However, there are always numerous domain-specific customizations required to model the context (as a general rule, the question and answer can be modeled using language features such as from a recurrent neural network). For example, Malinowski's work connecting words in questions to features in images uses a typical RNN-based model of questions and answers, but depends on a custom model for extracting those features from the images~\cite{malinowski2015ask}.  In short, the key difficulty in implementing QA systems boils down to: 1) obtaining a proper dataset, and 2) designing a suitable domain-specific model of the context.  How to design such a system for source code is an active research question.

In this paper, we present a QA system for answering ``basic'' programmer questions about subroutines in programs (the subroutines are the context about which questions are asked).  A ``basic'' question is a question about a small detail of a method, such as ``what are the parameters to the method convertWavToMp3?''  We use basic question because: 1) Eberhart~\emph{et al.}~\cite{eberhart2020apizateaser} found that programmers ask these questions about Java methods during actual programming tasks and isolated five types of these questions, and 2) the ``basic'' questions provide an excellent stepping-stone problem towards larger problems later.  We built question and answer templates and paraphrases based on Eberhart~\emph{et al.}'s basic question types to construct a dataset for 1.56m Java methods.  We then designed a QA system based on a neural encoder-decoder model.

We evaluated our work in two ways.  First, we used automated metrics over a large testing set of around 67k Java methods, to estimate how our approach would generalize.  Second, we performed an experiment with 20 human experts, to determine how well our model responds to actual human input for a subset of 100 randomly picked methods out of the 67k test set.  We explore evidence of \emph{how} our model learns to recognize pertinent facts in source code and generate English responses (in the spirit of explainable AI~\cite{samek2017explainable, ras2018explanation}).  

\section{Problem, Significance, Scope}
\label{sec:problem}

This paper is intended as a stepping stone towards a full-fledged QA system for software engineers.  Current technology in the NLP/AI domain is designed to answer questions from a given context, as described above.  The scope of this paper is to demonstrate how to adapt these technologies so that information about source code can be learned from large datasets, and that information used to answer questions about source code.  It is not reasonable to expect to build a QA system capable of answering arbitrary questions about programming in a single paper.  Existing peer-reviewed literature has not even demonstrated that context-based QA systems can learn from source code to interpret and answer natural language questions.  This paper is a proof-of-concept in that direction.


We understand that programmers probably would not use a QA system alone for basic informational questions about source code.  After all, the return type, parameter list, etc., of a function is readily available from reading the source code or summarizing documentation.  However, it is important to recognize that a QA system is usually not intended to be used on its own.  
Instead, a QA system for these questions is a key component in the big picture of conversational AI systems for programmers.  Robillard, with thirteen leading co-authors in the area of program comprehension, make the case clearly in a paper summarizing the outcomes of a relevant workshop in 2017~\cite{robillard2017demand}: they ``advocate for a new vision for satisfying the information needs of developers'' which they call \textit{on-demand developer documentation}.  The idea is that we as a research field should move towards machine responses to programmer information needs that are customized to that programmers' software context and individual questions.  But to get to that point, we (the research community) need to solve a few smaller problems that are currently barriers to continued progress.  This argument mirrors those made repeatedly in the AI research community generally~\cite{ward2016challenges, johnson2019no}, that smaller problems must be solved and used as a wedge against larger ones, towards the long-term goal of a meaningful conversational AI.

A QA system for basic programming information about subroutines is one of those wedge problems in program comprehension.  A successful system would not only answer the narrow problem at hand, but offer insights into issues of how to model and extract features from source code, how to interpret programmer information needs, and how to understand the vocabulary that programmers use that is different from general word use. Our paper demonstrates how a neural network can be used to recognize question type, find the answer in code and comments, and answer with a human readable grammatically correct format. Hence, the efficiency of this system is measured not only in accuracy of the information retrieved but the ability to recognize previously unseen human user questions and find as well as return correctly formatted answers from context. In the long run, our plan is to include this work as part of a larger interactive dialogue system for helping programmers read and understand source code. (\emph{Some citations omitted to comply with double-blind review policy.})

\vspace{-0.1cm}
\section{Background \& Related Work}
\label{sec:background}

This section covers background technologies and closely-related work in both NLP/AI and SE research venues.

\begin{figure}[b!]
	\vspace{-0.3cm}
	\centering
	\includegraphics[width=8cm]{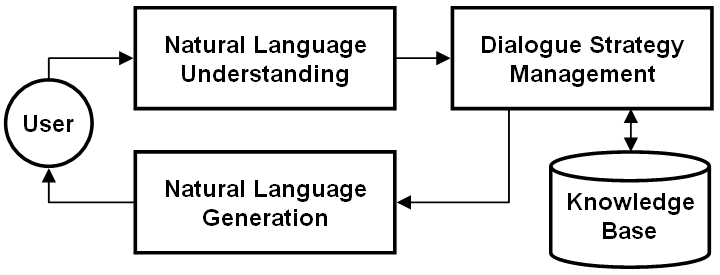}
	\vspace{-0.1cm}
	\caption{Stereotyped dialogue system described by Rieser and Lemon~\cite{rieser2011reinforcement}.  In this paper, the knowledge base consists of the source code of subroutines, while the understanding and generation components are learned via a neural net from a dataset we create.  We predefine the strategy based on experimental findings reported by Eberhart~\emph{et al.}~\cite{eberhart2020apizateaser}.}
	\label{fig:idsoverview}
\end{figure}

\vspace{-0.1cm}
\subsection{Interactive Dialogue Systems}
\label{sec:ids}

The anatomy of an interactive dialogue system is neatly articulated in a recent book by Rieser and Lemon~\cite{rieser2011reinforcement} and summarized in Figure~\ref{fig:idsoverview} below.  There are essentially four components.  First, a knowledge base is created to hold information relevant to the conversation, such as images about which questions are asked~\cite{malinowski2015ask,chen2019bidirectional}, or maps about which directions may be obtained~\cite{li2015gated, graves2016hybrid, xiong2016dynamic}, or restaurants which may be recommended~\cite{lemon2011learning}.  Second, a natural language understanding component is responsible for converting incoming text into an internal representation of what was said.  Often this starts with labeling the text with a dialogue act type~\cite{wood2018detecting, kumar2017dialogue, chen2018dialogue, blunsom2013recurrent, burger2000verbmobil} (e.g., as a question, a followup statement, a positive or negative comment).  But it also includes extracting relevant information necessary to form a response.  E.g., whether a user wants to know about the return type or parameter list of a subroutine.

The third component is dialogue strategy management.  This component decides how to respond as well as how to extract information necessary to make the response.  It uses the knowledge base to help make this decision and searches the knowledge base for information relevant to the response.  Note that the notion of ``strategy'' refers to the decision-making process that the machine follows, and is distinct from the natural language in the conversation~\cite{he2018decoupling}.  For example, if presented with a comment about the weather, some agents would respond with a summary of the predicted weather, some would respond with a suggestion to take an umbrella, while still others would ask a question about the user's preference for summer or fall.  But the decision about how to respond is not related to the words actually used to render a response.

Fourth, natural language generation techniques lie along a spectrum, one extreme of which is a templated, rule-based approach~\cite{Reiter:2000:BNL:331955} while the other extreme is a purely data-driven (usually deep learning-based) approach~\cite{deng2018deep}.  An example of a hybrid system is one in which canned responses are used to train a neural net (which allows more flexible combinations of the responses), or data-driven selection from a set of candidate template responses.  For a time, there was a belief that language understanding, strategy, and generation could be combined into a single module based on deep learning, but that belief is in strong decline for most applications~\cite{ward2016challenges, graves2016hybrid, he2018decoupling}.

QA systems fit into this anatomy of interactive dialogue systems in two ways.  First, a conversational system in an ongoing discussion with a user may include several subsystems to handle different situations, and pass control to a QA subsystem as needed.  Second, a QA system itself generally follows the same design.  The strategy component tends to be simpler than most systems due to the assumption that a single question from a user will elicit a single answer.  However, a QA system may need to cope with different types of questions and extract information from various artifacts -- both decisions that fall into the category of strategy.  In practice, the strategy tends to be encoded into the model via dataset design, rather than manual modification of the model.

Research into dialogue systems for software engineering is generally either foundational / dataset generation and analysis, or implementations of experimental dialogue systems.  Key foundational and dataset analysis work includes Maalej~\emph{et. al}~\cite{maalej2013patterns}, Eberhart~\emph{et al.}~\cite{eberhart2020apizateaser}, and several others~\cite{meng2018application, maalej2014comprehension, monperrus2012should, aghajani2018large, head2018not, robillard2011field, wood2018detecting}.  A recent survey discusses dialogue systems in SE~\cite{arnaoudova2015use}, which includes several experimental systems~\cite{tian2017apibot, ko2004designing, pruski2015tiqi, bradley2018context}.  These systems are related to this paper in the sense that they are prototypes of dialogue systems for SE problems, but are not directly comparable because they solve very specific problems and are based on scrutiny of highly-specialized domain knowledge.  While one may perform quite well in one situation, it may be unsuitable for others; for instance, OpenAPI Bot \cite{openAPI} is capable of answering programmer questions about API components, but requires well-documented API specifications in a particular format. A robust QA system that relies on AI may be better able to generalize to a broader range of software and documentation formats by automatically identifying relevant features in development artifacts.  This specialization is typical of dialogue systems in all domains~\cite{rieser2011reinforcement}, so the way to evaluate an approach is to compare implementation alternatives rather than different dialogue systems~\cite{beringer2002promise, walker1997evaluating}.

\vspace{-0.1cm}
\subsection{Neural Encoder-Decoder}
\label{sec:encdec}

The encoder-decoder model is the current state-of-the-art for QA systems, as described in several surveys~\cite{deng2018deep, chen2017survey, gao2019neural, chen2019graphflow}.  To pick one very recent and related paper that exemplifies how dialogue systems based on the encoder-decoder model work, consider Lin~\emph{et al.}~\cite{lin2019task}.  The paper presents a memory model to augment the encoder of a typical encoder-decoder design, and compares it to other encoder-decoder models over publicly-available datasets.  This paper is similar, but instead of a model tuned for general conversations, we propose an encoder model for this specific SE problem to gain domain specific knowledge.

The encoder-decoder design has been clearly described in several papers, and we discuss details in our approach section.  In general, the design includes an encoder, which receives a natural language input from the user and the knowledge base.  The encoder outputs a vector representation of the input, usually via a recurrent neural network (RNN).  The decoder receives the example desired output during training, and learns to generate that output from the corresponding input.  During inference, the model outputs one word at a time, and uses the output predicted ``so far'' to help predict the next word.

The encoder-decoder design ballooned in popularity after Bahdanau~\emph{et al.}~\cite{bahdanau2014neural} introduced an ``attentional'' variant that allows the decoder's vector representation to focus on sections of the encoder's representation during training, i.e. to create a dictionary of words in one language in the decoder to another language in the encoder.  Specific designs such as the famed seq2seq model have motivated thousands of papers, well beyond what we can describe in this section.  Thus we direct readers to several surveys~\cite{young2018recent, shrestha2019review, pouyanfar2018survey, yu2020crossing}.  Within software engineering literature, the encoder-decoder design is seeing increased use for tasks such as code completion~\cite{hayati2018retrieval}, code summarization~\cite{leclair2019neural, leclair2020improved}, and automated repair~\cite{chen2019sequencer}.

\vspace{-0.1cm}
\section{Approach}

Our approach aligns with the related work described in the previous section: the overall architecture is based on the dialogue system design in Figure~\ref{fig:idsoverview}, and the implementation is based on a neural encoder-decoder model.  The key novelty in the model is the representation of the knowledge base.  The key novelty in the overall architecture is the crafting of our dataset to train the neural model.  These set up a novel evaluation, that shows how these models work in a QA system for program comprehension of functions.  In the long run, we plan for this QA system to be a component of a much larger dialogue agent to be designed in a future paper.  An overview of the components of our dialogue system follows:

\textbf{Dialogue Strategy Management}  Recall that dialogue strategy management involves decisions both on 1) how to respond, and 2) how to extract the information necessary to make a response.  For (1), we craft a dataset that includes eight types of questions that we found in recently-released simulation experiments with programmers.  While those experiments were performed by another study, we completed the analysis of the eight questions for this paper.  The dataset design represents our manual effort in designing the strategy the system should follow, but the strategy itself will be learned during training and encoded in a neural model.  For (2), we use an attention mechanism in our neural model between the input question and the knowledge base, to learn which components of the knowledge base pertain to specific questions.  Details of our dataset design are in Section~\ref{sec:dataset}.  Details of the attention mechanism and neural model are in Section~\ref{sec:impl}.

\textbf{Knowledge Base}  The knowledge base consists of the source code of the subroutines.  We use a collection of Java methods provided by Linstead~\emph{et al.}~\cite{Linstead2009} and further processed by LeClair~\emph{et al.}~\cite{leclair2019recommendations}.  In total, the knowledge base includes 2.1m Java methods from over 10k projects. These java methods are filtered for non-tokenized characters and fed as raw source code and questions-answer pairs to a 3 input seq2seq model. 

\textbf{Natural Language Understanding / Generation}  We use recurrent neural networks with word embedding vector spaces to implement the encoder and decoder.  The encoder is essentially the component that implements the natural language understanding, and the decoder implements the language generation.  This structure is closely in line with a vast majority of recent data-driven QA systems (see Section~\ref{sec:encdec}).

\vspace{-0.2cm}
\subsection{Dataset Preparation}
\label{sec:dataset}
\vspace{-0.1cm}

We prepared a dataset to train the neural model described in section 4.3.  This section describes how we structure our dataset to represent knowledge about how programmers ask questions and how to respond.  Note that we do not explicitly write rules into our dialogue strategy management; this dataset contains those rules implicitly, from which the neural model learns later.  To be clear, we do not merely feed the network all data collected empirically and expect the model to learn proper behavior, and justify overall posture towards dataset design: the decisions for creating the dataset \emph{are} the decisions that will be encoded as dialogue strategy management.

We build the rules for generating our dataset based on empirical data made available to us pre-release.  Eberhart~\emph{et al.}~\cite{eberhart2020apizateaser} conducted an experiment in which 30 programmers solved programming challenges with the help of a simulated interactive dialogue agent (a ``Wizard of Oz'' study design).  The authors then annotated each question with one of twelve types of API information needs (these API info needs were determined in an earlier TSE paper by Maalej and Robillard~\cite{maalej2013patterns}).  Eberhart~\emph{et al.} found that over 90\% of questions had three information needs: {\small \texttt{functionality}}, {\small \texttt{patterns}}, or {\small \texttt{basic}}.  In the long run, a dialogue agent will need to handle all of these.  But the scope of that challenge is far too much for one paper. As an early attempt at the problem, we focus on {\small \texttt{basic}} questions which tend to be more self-contained, have concrete single-turn answers, and are overall easier to answer.

A {\small \texttt{basic}} question is one in which a programmer asks for key information about the components of code.  The ``components'' were almost always subroutines rather than classes etc.  The ``key information'' included things like the return type, the function parameters, or a high level description (such as a summary comment from JavaDocs).  Approximately 20\% of the questions asked by programmers in the study by Eberhart~\emph{et al.}~\cite{eberhart2020apizateaser} were {\small \texttt{basic}} questions.

We (independent of the analysis by Eberhart~\emph{et al.}) examined questions that were labeled {\small \texttt{basic}}.  The authors created eight categories of {\small \texttt{basic}} question.  The procedure was an open coding process in which the authors labeled each question with a specific information need from a subroutine. The authors worked together to resolve disagreements, rather than work independently and compute an agreement metric, in order to improve reliability of the data\footnote{Agreement metrics quantify reliability, but do not resolve disagreements.  Because we ultimately had to make decisions to create a dataset, we elected to resolve disagreements at the cost of a reliability metric, as suggested by Craggs and McGee~\cite{craggs2005evaluating}.}.

In the end we had eight types of {\small \texttt{basic}} question.  An important distinction is that six of the questions involved \textbf{known} subroutines i.e. the programmer already knew the correct method  For example, what the return type of method {\small \texttt{X}} is.  Three of the questions involved \textbf{unknown} subroutines i.e. the programmer did not know the correct method.  For example, which method takes an int as a parameter and returns a string.  We call questions with a known subroutine ``type K'' and questions with an unknown subroutine ``type U.'' 

\textbf{Type K questions (subroutine known):}

\begin{enumerate}
	\item What is the return type of \emph{method}?
	\item What are the parameters of \emph{method}?
	\item Give me the definition of \emph{method}.
	\item What is the signature of \emph{method}.
	\item What does \emph{method} do?
	\item Can \emph{method}, \emph{short task description}?
\end{enumerate}

\textbf{Type U questions (subroutine unknown):}

\begin{enumerate}
	\setcounter{enumi}{6}
	\item How do I \emph{short task description}?
	\item What method uses parameter type \emph{P} and returns type \emph{R}?
\end{enumerate}

\textbf{The scope of our QA system only includes type K questions.}  Type K questions involve a question, answer, and known context, which is in line with what QA models in NLP are equipped to solve (though, those models have not been adapted to source code).  Type U questions involve a search process for the correct subroutine, which would include code search and even dialogue between machine and programmer to decide on the correct subroutine.  These search tasks are extensive research problems with standalone research papers.  Therefore, we confine ourselves to the problem of answering {\small \texttt{basic}} questions about known subroutines.  Integrating code search, grounding dialogue, etc., is an area for our future work, to build on this paper.


\vspace{-0.2cm}
\subsection{Dataset Generation}
\label{sec:build}
\vspace{-0.1cm}

The next step is to generate a dataset with the isolated question types.  To do so, we obtained a large repository of Java methods, then generated example questions and answers for each question type using heuristics to automatically extract information from the methods.  The repository of Java methods is a set of 2.1m methods already filtered for duplicates and other errors, and paired with summary descriptions, provided at NAACL'19~\cite{leclair2019recommendations}. We further filtered out methods with duplicate and non-descriptive comments, leaving 1.5m methods.

To generate text for questions (1-4)  we extracted information from raw code for each method e.g. the return type.  For question (5), we used the summary description formatted grammatically. For question type (6), we used the summary description and method name in the question and the answer was simply ``yes'' or ``no.''  For question type (6) only, we included a positive and a negative example to maintain a balanced dataset.  This negative example consisted of a random summary description from another method (of a different name, to avoid picking an overloaded method name) in the same project paired with the method.  



To limit the vocabulary size, we replaced some information with tokens that direct the output interface to copy the information from the context directly, rather than learn to predict the information as part of the model. In particular, we use a  $<$$funcode$$>$  token to stand in for a method's definition, rather than require the system to copy method code one token at a time. This allows the user to have the same experience while reducing the vocabulary that the model has to learn.

The final step was to paraphrase each question and answer.  The Type K questions listed in Section 4.1 are archetypal examples of user queries expressing specific information needs; in practice, a programmer would not use exactly that language when forming a query.  To account for this variability, we wrote 15-25 paraphrased layouts of each question, and randomly chose one of them when generating questions and answers. In lieu of real programmer-helper QA dataset, we aim for the neural network to learn from paraphrases enough to recognize most, if not all the variations of these questions asked by the participants.  All replacements and paraphrases are available via our online appendix (see Section~\ref{sec:repro}).

To summarize, our procedure is:


\begin{algorithm}
\vspace{-0.3cm}
\SetAlgoNoLine
\For{each of the 1.56m methods M}
{
	\For{each question type T}
	{
		1. randomly select paraphrase template for T	\\
		2. generate question and answer using template	\\
		3. preprocess code of M to serve as context		\\
		4. create 3-tuple: (question, answer, context)	\\
		\If{T == 6}
		{
			5. randomly select summary of different method \\
			6. create 3-tuple: (question, ``no'', context) \\ 
		}
	}
}
\vspace{-0.3cm}	
\end{algorithm}


The result of our dataset generation is a set of 10.88 million 3-tuples.  Each 3-tuple contains a question, an answer, and a context Java method.  For each of the 1.56m Java methods, we generated 7 type K questions and answers (one for question types 1-5, two for question 6).  To best improve reproducibility, we maintained the training/validation/test - 90/5/5 splits that ensure no contamination between projects or methods, provided by LeClair~\emph{et al.}~\cite{leclair2019recommendations}.

\vspace{-0.1cm}
\subsection{Neural Model}
\label{sec:impl}

\textbf{Why Neural Networks?} Neural models enable flexible natural language understanding and generation in few steps, without the need for manually-written rules to extract information from context.  A traditional alternative to a neural model is a simple approach based on classification of incoming questions and rules to extract information.  However, this seemingly-obvious alternative is not in line with recent work from the NLP research area for context-based Q/A systems.  As Wiese~\emph{et al.}~\cite{wiese2017neural} point out, recent advances in neural models have led to ``impressive performance gains over more traditional systems.''  

In contrast, our model falls clearly in line with related work from the NLP research area on context-based Q/A systems (see Section~\ref{sec:encdec}). From an ML perspective, one novel aspect to this paper is that we show how the neural model can learn features in the source code when given only that code as a context, and questions/answers about the context.  This is important novelty, along the lines of Wiese~\emph{et al.}~\cite{wiese2017neural} when they showed how neural Q/A models can learn from biomedical text data versus other highly specific areas e.g. technical support conversations~\cite{castelli2019techqa} or even religious texts~\cite{zhao2018finding}.  The point is that domain adaptations are considered important contributions and are not merely applying technology X to data Y. Neural networks add flexibility and variance to both the inputs they can handle in a study like ours as well as the responses that will be evaluated subjectively by  programmers.

\textbf{Overview} At a high level, our neural model is similar to context-based Q/A systems described in related work (see Section~\ref{sec:encdec}).  These systems use a question and context as encoder input and an answer as decoder input (for training).  The model learns to predict answers one word at a time.  Our model follows this same basic structure; the question and answer are generated for each function, and the context is the source code of the function.

\begin{figure}[b!]
	\vspace{-0.9cm}
	\centering
	\includegraphics[width=8cm]{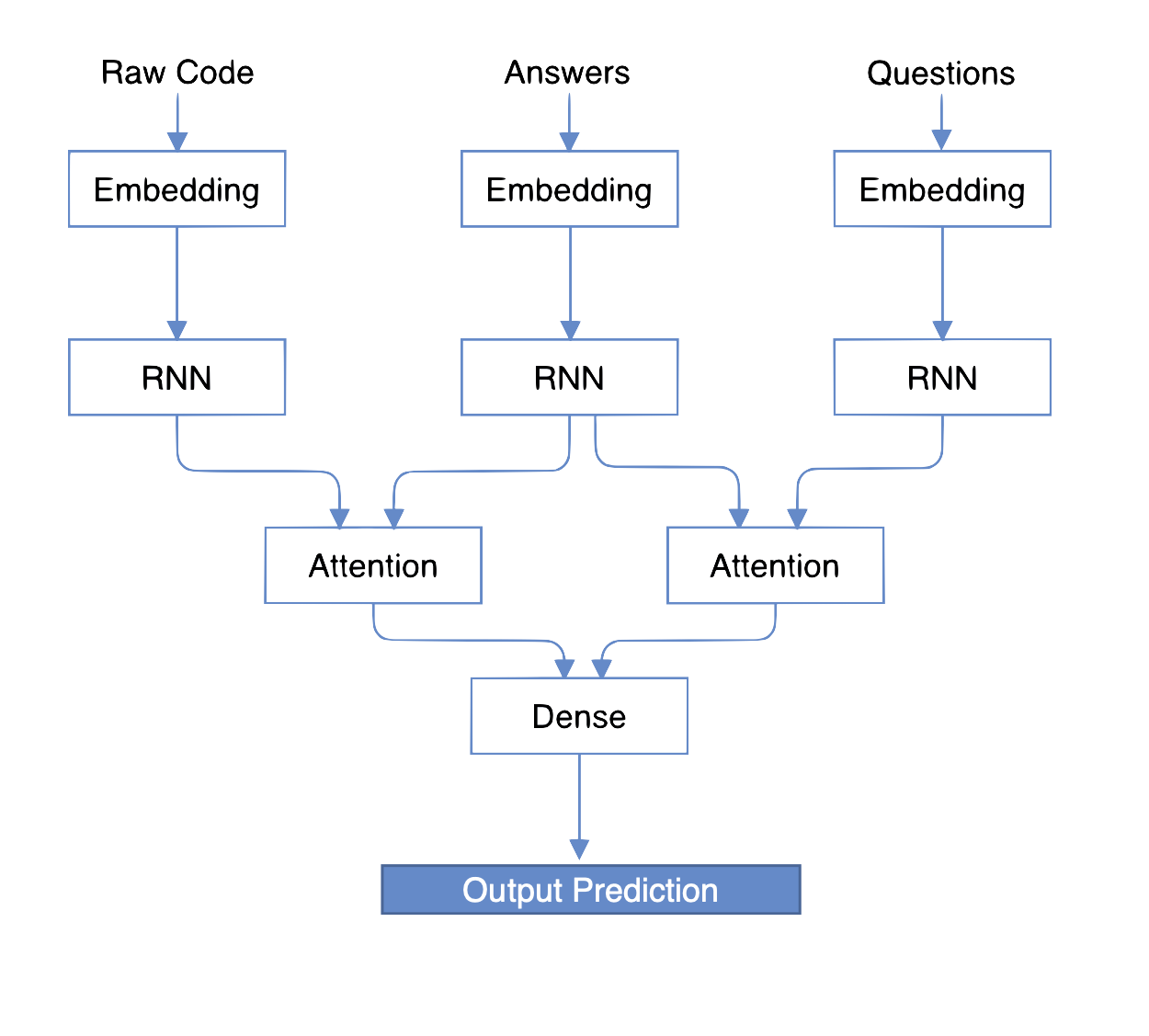}
	\vspace{-0.7cm}
	\caption{The neural model and data pipeline.}
	\label{fig:flowchart}
\end{figure}

Our approach is based on the model released by LeClair~\emph{et al.}~\cite{leclair2019neural} at ICSE 2019.  We chose that model because: 1) it was designed to accommodate source code as input instead of only text, 2) a thorough reproducibility package is available, and 3) it does not rely on preprocessing of data into semantic formats or graphs like other state of the art models such as \cite{code2seq}, \cite{code2vec}, \cite{multimodal}. That model was designed to generate natural language descriptions of source code (so-called ``source code summarization'').  The inputs to the model's encoder were preprocessed source code, a flattened abstract syntax tree.  The input for training for the decoder was the example summary.

Our modifications, in brief, are to make the model's encoder inputs the raw source code (not preprocessed), to add an input for the user query/question to the encoder, and to change the decoder's training input to example answers to the questions.  We used raw source code instead of preprocessed source code because we are interested in the model's ability to learn where code features are such as the return type, parameters, etc., unlike LeClair~\emph{et al.} who were more interested in extracting text features such as identifier names. Their preprocessing steps removed information that we found to be critical in helping the model learn features about code. This is critical for future applications of a VA to fetch information from any raw source code found online without delay or dependencies.

\textbf{Details} We provide the Keras implementation of our model to maximize clarity and reproducibility, following the example of LeClair~\emph{et al.}~\cite{leclair2019neural}, via our online appendix (Section~\ref{sec:repro}).  The relevant model code is in file {\small \texttt{qamodel.py}}.

\vspace{-0.1cm}
\subsection{Training Procedures}
\label{sec:training}

Our training procedure is based on the ``teacher forcing'' technique~\cite{lamb2016professor, logar1993comparison, doya2003recurrent} in which the model receives only correct examples from the training set and is not exposed to its own errors.  Recall that an encoder-decoder architecture typically predicts output sequences one item at a time.  For example, given a question ``what is the return type of function X?'', the model would generate an answer by predicting the first word of the answer:

\vspace{-0.05cm}
{\footnotesize
	\begin{verbatim}
	[ question ] + [ code ]
    => [ "the" ]
	\end{verbatim}
}
\vspace{-0.05cm}

Then it would use the first word prediction as a new input to the decoder, to predict the second word, and so on:

{\footnotesize
	\begin{verbatim}
	[ question ] + [ code ] + [ "the" ]
    => [ "method" ]
	[ question ] + [ code ] + [ "the method" ]
    => [ "returns" ]
	[ question ] + [ code ] + [ "the method returns" ]
    => [ "a" ]
	[ question ] + [ code ] + [ "the method returns a" ]
    => [ "long" ]
	\end{verbatim}
}


Yet this is how the model behaves during inference.  To train the model, following the teacher forcing procedure, we provide the model each example one word at a time.  So, in the above example, we would provide the model with ``the'' followed by the reference output ``method'', then ``the method'' with the reference output ``returns'', and so on.  If the model makes an incorrect prediction, we use back propagation to correct the model, and then substitute the correct reference output for the next step -- the model is not permitted to use its own erroneous prediction as the next input.  



\section{Evaluation}
\label{sec:eval}

We conduct an experiment with human users to evaluate our QA system.  Note that our ultimate intent for this QA system is to serve as a component of a much larger conversational AI (see Sections~\ref{sec:problem} and~\ref{sec:ids}).  Therefore, our experimental setup is a controlled environment in which we test specific inputs and outputs generated by human users.  We are \emph{not} attempting to evaluate the system ``in the wild'' because the system is not intended to be used standalone, and because the larger conversational AI system does not yet exist.

\vspace{-0.2cm}
\subsection{Research Questions}
\label{sec:rqs}

Our research objective is to determine the degree to which our QA system is able to answer the eight questions about subroutines we determined in Section~\ref{sec:dataset}.  We ask the following Research Questions (RQs) towards this objective:

\begin{description}
	\item[RQ$_{1}$] How well does our QA system perform in terms of relevance, accuracy, completeness, and conciseness?
	
	\vspace{0.05cm}
	
	\item[RQ$_{2}$] How does the performance vary across the six question types for which we designed the system?
	
	\vspace{0.05cm}

	\item[RQ$_{3}$] What features in the context are the most important for the model to use when answering a question?
\end{description}

The rationale behind RQ$_1$ is that good responses by any QA system should score well across these degrees: relevance, accuracy, completeness, and conciseness.  Relevance, to observe if the model recognized the question and tried to get requested information For example, if asked Q1 and the return type is int, but the answer given by the model is ``The return type is double'', this is inaccurate but relevant. Accuracy, because independent of any other factors the response should not contain false information.  Completeness, because responses should contain \emph{all} information needed to answer the question.  Conciseness, because responses should contain only the necessary information needed. All participants were given examples of situations where it may be confusing to rate these scores to particularly differentiate between relevance and accuracy. We derived these four degrees of text generation quality from related SE literature on code description generation~\cite{sridhara2011automatically, mcburney2016automatic}.  The rationale behind RQ$_2$ is that the system may perform well for some questions but not others.  In particular, it may perform well at extracting information such as the return type of a subroutine, but struggle for other questions such as returning a description of a subroutine.  We ask RQ$_3$ for insights into the model's behavior.  Neural models tend to be highly effective for text comprehension and generation tasks, but are notorious for producing black box responses that are difficult to understand.  


\vspace{-0.2cm}
\subsection{Methodology}
\label{sec:methodology}

\begin{figure}[b!]
	\vspace{-0.3cm}
	\centering
	\includegraphics[width=8cm, height=5cm]{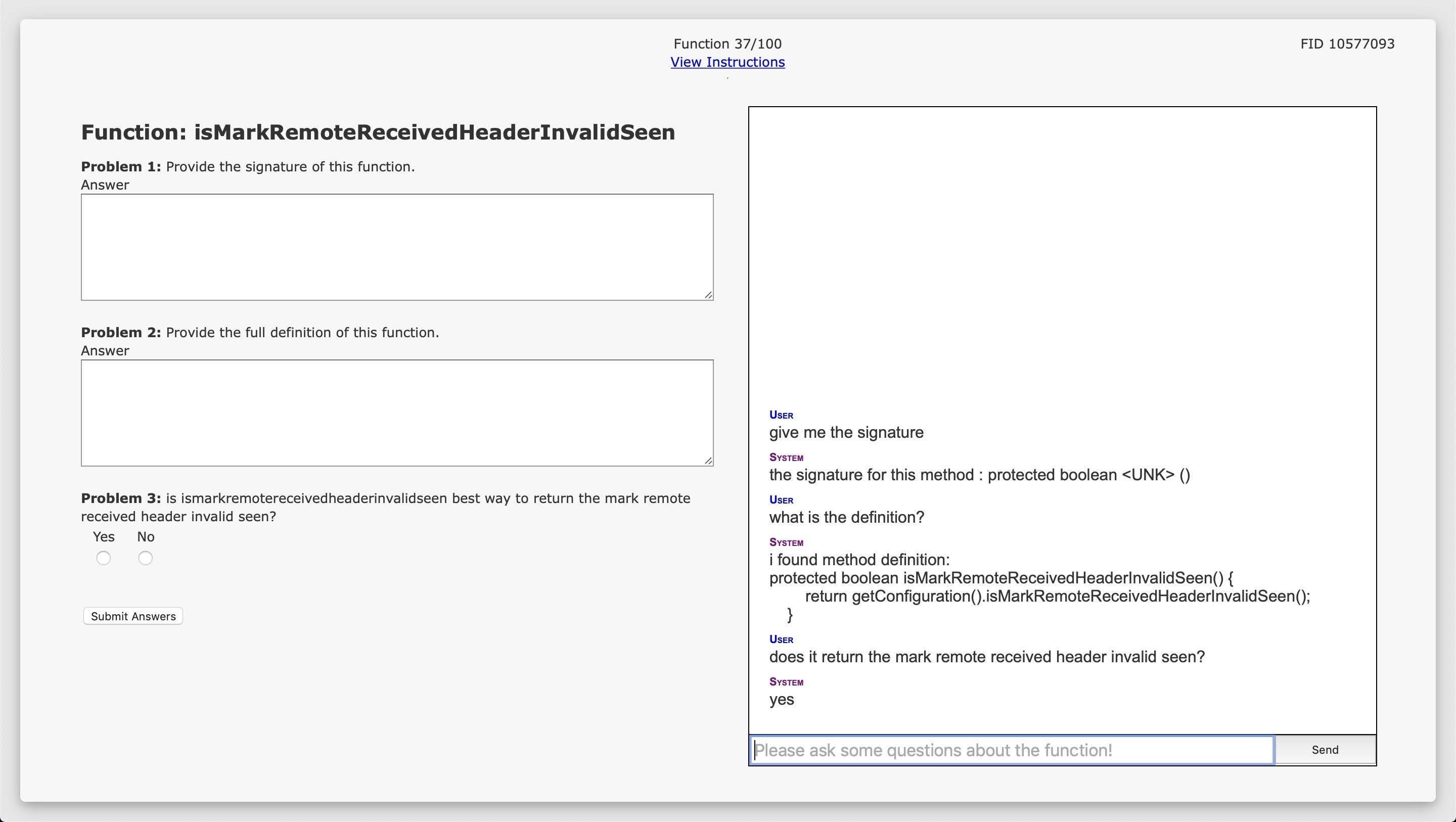}
	\vspace{-0.1cm}
	\caption{The interface that programmers used to communicate with the QA system during our experiment.}
	\label{fig:interface}
\end{figure}

Our methodology for answering RQ$_1$ and RQ$_2$ is to conduct a user study in which human programmers ask questions to a QA system and evaluate the responses.  To limit the scope of the experiment, we control the study conditions so that the programmers only ask questions related to the six Type K questions highlighted in Section~\ref{sec:dataset}.  We recruited professional programmers from United States and Europe via an online job platform (demographics of study population are in the next section).  We also created a web interface for the programmers to communicate with the QA system.  A screenshot of this interface is in Figure~\ref{fig:interface}.  The interface then prompts the programmers to rate the responses on a 0-4 scale ranging from Strongly Agree, Agree, Neutral, Disagree, or Strongly Disagree for the quality prompts shown in Table~\ref{tab:quality}.

\begin{table}[t]
	\begin{center}
		\caption{Quality prompts for the user study, corresponding to the quality criteria (relevance, accuracy, completeness, and conciseness). The responses for P1-P4 are ``Strongly Agree'', ``Agree'', ``Neutral'', ``Disagree'', and ``Strongly Disagree.''  
		\label{tab:quality}}
		\begin{tabular}{p{0.3cm}|m{7cm}}
			\hline
			$P_{1}$&\vspace{0.07cm}Independent of other factors, the response is relevant to my question, even if the information it contains is inaccurate. \vspace{0.15cm} \\
			$P_{2}$&The response is accurate, even if it is not relevant to my question. \vspace{0.15cm} \\
			$P_{3}$&The response is missing important information, and that can hinder my understanding. \vspace{0.15cm} \\
			$P_{4}$&The response contains a lot of unnecessary information. \\ \hline
			$P_{5}$&\vspace{0.07cm}Do you have any general comments about the response? \\
			\hline
		\end{tabular}
	\end{center}
	\vspace{-0.4cm}
\end{table}

\textbf{Rationale} Our study design is similar to previous experiments by Sridhara~\emph{et al.}~\cite{sridhara2011automatically} and McBurney~\emph{et al.}~\cite{mcburney2016automatic}.  We used similar wording of our prompts to study participants and the same four options.  The only difference we made was to add another option for Neutral in case the model returns a nonsensical reply (which can happen for our neural model but was very unlikely in the templated systems of code comment generation in those previous studies).  We added the Neutral option as a middle ground, to avoid forcing participants to make decisions on possible nonsensical responses.

Another similarity is that we find ourselves in the same situation as Sridhara~\emph{et al.}~\cite{sridhara2010towards} in their ASE paper: no baseline exists for comparison.  Even cross domain studies that could be used as baseline use real data, hence they are ~\emph{not} comparable to our study with a synthesized dataset. To our knowledge, no QA system has been designed to answer these specific questions in a natural language format.  Different tools do exist for some questions.  For example, question (5) could be thought of as a code summarization question, while questions (1-4) could be answered by just reading the subroutine itself.  Yet recall that we are not seeking an ``in the wild'' evaluation -- we need to evaluate the input and output of the model in situ with the natural language understanding and generation components of the approach.  Therefore, we follow the example of these earlier papers and focus on a deeper analysis of the responses across multiple quality criteria, instead of comparing metrics across competing approaches (since they do not yet exist).

Note also that we do not use BLEU scores or other automated metrics.  A human study is vital for two reasons.  First, we need to evaluate specific subjective qualities rather than similarity to a ground truth.  Second, the ground truth in our dataset (i.e. the answer component of the question, context, answer tuples, see Section~\ref{sec:dataset}) is generated by us.  We use it as training data, but it would not be appropriate to use as testing data since it would include our own biases.   

\textbf{Experiment Procedure} In the experiment, we gave each programmer a ``quiz'' to fill out with the assistance of the QA system (see Figure~\ref{fig:interface}).  Each page of the quiz gave the name of a particular Java method.  Only the method name was shown, not the method body. For each method, three Type K questions (see Section \ref{sec:dataset}) were chosen randomly. Below the method name, there were three prompts, derived from the chosen Type K questions. We phrased the prompts as imperative statements (e.g. ``Provide the return type of this function'') to avoid priming the programmers with a particular question format. We instructed programmers to use their own words to ask for information from the QA system. We asked programmers not to copy questions, but we allowed them to copy answers from the QA system for the quiz.  A programmer could ask the QA system as many queries as he or she wanted.

After answering the question prompts for a particular method, programmers were brought to a new page that asked them to rate each of their interactions with the QA system for that method. For each interaction (consisting of a user query and the QA system's response), we asked the programmers to answer the five quality prompts listed in Table \ref{tab:quality}. When they were done, they could press a button to bring up the next method, and a new set of prompts.

In short, we used a quiz format to encourage programmers to ask the QA system certain types of questions n their own words.  They then rated the responses using the quality prompts, so we could determine whether they obtained the correct information in the end.  Materials relating to the quiz design may be found in our online appendix (Section~\ref{sec:repro}).

For clarity in the experimental results section, we use the following vocabulary: 1) a ``question'', Q1-6, is one of the six Type K question types we use in our experiment and described in Section~\ref{sec:dataset}, 2) a ``query'' is text typed by the user into the experiment interface until the return key, hitting the return key triggers the interface to send the text to the prediction model and receive an answer back, and 3) a ``quality prompt'', P1-5, is one of the requests we make of users to rate the model's answer.  The users see three questions per function.  They may write as many queries as they wish.  Then they respond to five quality prompts for the answer to each query.

\subsection{Participants}
\label{sec:participants}

We recruited 20 participants for our experiment. These participants had professional experience ranging from three to 15 years.  Programmers were compensated at a flat rate of US\$60/hr, market rate in our region, regardless of performance speed.  Each programmer spent a total of 40 minutes on as many quiz pages as they could.

\subsection{Subject Java Methods}
\label{sec:subjectmethods}

We used a total of 100 Java methods in our experiment.  We sourced these methods from the test split -- the model had not seen them during training.  We picked randomly so that no programmer saw the same method more than once and each method was shown to at least three programmers.  But given the vicissitudes of any study with humans (fatigue, differing speeds), not all methods ended with three ratings.

\subsection{Threats to Validity}
\label{sec:threats}

Like any paper, this experiment carries threats to the validity of its conclusions.  One threat is mainly external as defined by Yin~\cite{yin2017case} and not internal or case reliability due to uniform process and objective analysis using a neural net.  The number of functions that we could ask any one person to evaluate is limited.  We chose a random selection from a large, curated dataset ensuring each function was seen by more than one person, but it is still possible that a different selection would result in a different result.  Similarly, a different set of programmers might give different answers.  We attempted to mitigate this risk by asking over 20 participants.  

One threat to construct validity ~\cite{yin2017case} is the experimental instructions explained by the administrator, which we attempt to mitigate by using a scripted description by the single administrator.  Technical reliability is a risk outside of study design, we mitigate varying results from the model itself by ensuring consistent random seeds and experimental conditions (available via our online appendix), however random factors in GPU performance or software could lead to different results.


\section{Experimental Results}
\label{sec:results}

This section describes the results of the experiment: our answers to our RQs and supporting evidence.
\begin{table}[hb!]
	\caption{Sequential interaction of one participant.We noticed incorrect response with out-of-vocab words like ``what's''}
	\label{tab:interactions}
	\vspace{-0.2cm}
	\begin{tabular}{lm{6cm}}
		\emph{user query}   & what does it return? \\
		\emph{model output} & the return type for this method is void\\
		\emph{user query}	& does it set the domain name?\\
		\emph{model output} & yes\\
		\emph{user query}	& what's its purpose?\\
		\emph{model output}	& no\\
		\emph{user query}	& what does it do?\\
		\emph{model output}	& this method sets the domain name\\
	\end{tabular}
	\vspace{-0.4cm}
\end{table}

\vspace{-0.1cm}
\subsection{RQ$_1$: Overall Performance}

\begin{table}[b!]
	\vspace{-0.3cm}
	\caption{Performance statistics of participants in the experiment.  Each participant worked for 40 minutes.  We asked three questions for each method.  Participants worked at their own speeds and were allowed to ask multiple queries.}
	\label{tab:speedstats}
	\vspace{-0.2cm}
	\begin{tabular}{llll}
		& Mean	& Min	& Max	\\
		Methods Evaluated per Participant     &	19      & 7		& 38	\\
		Queries per Participant               & 70  	& 37	& 117	\\
		Queries per Method                    & 3.8		& 2		& 10	\\
		Queries until ``Correct'' Response    & 1.2		& 1		& 8		\\ 
	\end{tabular}
	\vspace{-0.4cm}
\end{table}

\begin{figure*}[!tp]
	\centering
	\begin{subfigure}{7cm}
		\includegraphics[clip,width=7cm]{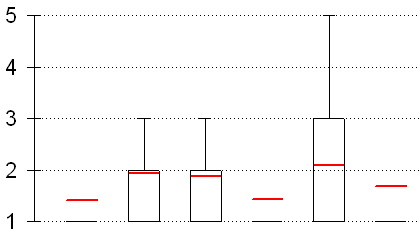}
		\vspace{-0.6cm}
		\caption{P$_1$ Relevance}
	\end{subfigure} \hspace{1cm}
	\begin{subfigure}{7cm}
		\includegraphics[clip,width=7cm]{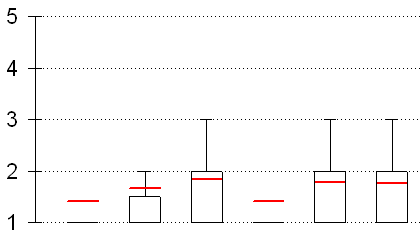}
		\vspace{-0.6cm}
		\caption{P$_2$ Accuracy}
	\end{subfigure}
	
	
	\begin{subfigure}{7cm}
		\includegraphics[clip,width=7cm]{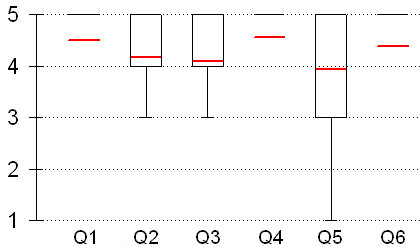}
		\vspace{-0.5cm}
		\caption{P$_3$ Completeness}
	\end{subfigure} \hspace{1cm}
	\begin{subfigure}{7cm}
		\includegraphics[clip,width=7cm]{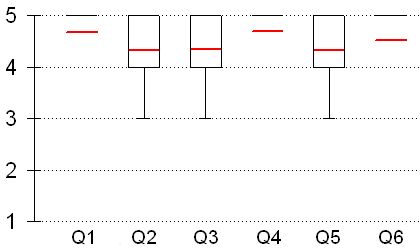}
		\vspace{-0.5cm}
		\caption{P$_4$ Concision}
	\end{subfigure}
	
	\vspace{-0.1cm}
	\caption{Boxplots of answers to quality prompts (relevance, accuracy, completeness, and concision) for each of the question types from Section~\ref{sec:dataset}. For P1 and P2 lower score is better however, for P3 and P4 higher scores are better as those questions are posed negatively.  Performance is highest for Q1 and Q4 and worst for Q5: ratings of relevance and completeness tend to be worse for Q5 than for other question types.}
	\label{fig:rq2_boxplots}
	\vspace{-0.3cm}
\end{figure*}

\begin{figure}[!b]
	\centering
	\vspace{-0.3cm}
	\includegraphics[width=7cm]{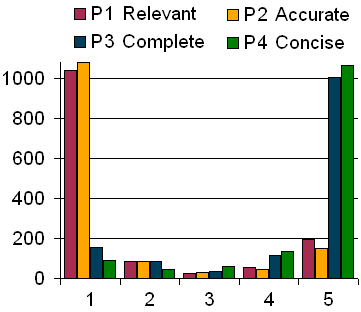}
	\vspace{-0.1cm}
	\caption{Histograms of the user responses to the quality prompts in Table~\ref{tab:quality}.  Recall that P1 and P2 are asked in a positive tone (so 1-2 scores are better) while P3 and P4 are in a negative tone (so 4-5 are better).  Participants tended to find the model's responses to be of good quality.}
	\label{fig:rq1_histogram}
\end{figure}

Overall, the model tends to generate reasonable responses. Table~\ref{tab:interactions} shows an interaction, the model responds correctly when words are in vocabulary but for out-of-vocab query. Figure~\ref{fig:rq1_histogram} gives an overview.  The figure is a histogram of all user answers to the quality prompts from Table~\ref{tab:quality}.  Recall that 1=``Strongly Agree'', 2=``Agree'', 3=``Neutral'', 4=``Disagree'', and 5=``Strongly Disagree'' to the prompt text.  Prompts 1 and 2 are worded positively (so agreement is better) while prompts 3 and 4 are worded negatively (so disagreement is better).  For example, for P$_1$ about relevance, a vast majority of responses received a score of Strongly Agree or Agree.  For P$_3$, a vast majority of responses indicated disagreement i.e. important information was \textit{not} missing.  Also note that only a few of responses were rated as neutral; in general, responses were clear enough for participants to form an opinion. Most of the responses rated as neutral were gibberish model output. 

Two caveats should be understood.  First, different participants worked at different rates, so some participants are represented more in the data than others.  Table~\ref{tab:speedstats} quantifies these differences. Most participants evaluated between 15 and 20 methods, but there were a few outliers as is natural in samples of human populations (mean speed of 19 methods per 40 minute study is about 2 minutes per method, while 38 is a rate of about 1 minute/method).  Nonetheless we found the number of queries required to answer each question to be quite stable, with one query usually sufficing and two or more queries being quite rare.  The time required by each participant was dependent on their reading, typing and rating speed, more than the number of queries asked.

Second, the responses to each quality prompt are independent of other prompts.  So it is possible that a response receives a good score for P$_2$ and a poor score for P$_1$ (see Section 5.1).  To study this caveat, we derived a binary metric we call ``correctness'' by combining P$_1$ and P$_2$ scores.  A response receives a 1 if and only if both P$_1$ and P$_2$ scores are [1,2] -- i.e. a response is only ``correct'' if the participant strongly agrees or agrees that it is both relevant and accurate.  We found that 79\% of responses were ``correct'' and that it usually took only one query to receive a correct response.

We found that a key factor in the 21\% of incorrect responses to be the vocabulary size.  As mentioned in Section~\ref{sec:impl}, GPU memory limitations restrict both the input and output vocab size, despite our attempts to extend these by using GPUs with 16gb VRAM and low training batch sizes.  This limit affected our results.  A vast majority of the responses that were relevant but not accurate were ones with UNK tokens in the answer( i.e. model could find the answer correctly but the UNK token was an important part of the answer). Likewise, responses that were accurate but not relevant almost always had UNK tokens in the question (i.e. the participant wrote a query with out-of-vocab words in it) -- these UNK tokens likely caused the model to misunderstand the question and give an accurate response that was nonetheless irrelevant such as in Table~\ref{tab:interactions}. As with all human studies, unexpected or incorrect queries i.e. human error can also drive results as we allowed participants to interact with the models relatively freely after short guidelines.

\vspace{-0.1cm}
\subsection{RQ$_2$: Variation among Question Types}

We observe a small degree of variation among the question types in our experiment.  Recall from Section~\ref{sec:dataset} that we have a variety of question templates that we derived from six different question types corresponding to six key information needs programmers have.  (In the experiment, we confined participants to these information needs, but we had no restriction on the language that they could use to render a question.)  Recall that the rationale of this RQ is that the model may be better at understanding some information needs than others. 

Figure~\ref{fig:rq2_boxplots} contains boxplots of the answers for each quality prompt, divided across each question type.  For example, column Q2 of Figure~\ref{fig:rq2_boxplots}(a) shows that for Question 2, the mean of all responses to queries is about 2 (the red line), the interquartile range is 1 to 2.  The way to interpret this is that, among all queries written for Q2, participants either Strongly Agreed or Agreed that the query was relevant about half of the time.  Note that outliers are excluded for readability, but we did have at least one instance of each score.

In general, the model performs well for Q1 and Q4.  For both, the responses are dominated by optimal scores (1 for relevance and completeness, 5 for accuracy and concision).  This result implies that the model is successfully learning to recognize when participants were asking for those information needs, and also learns to extract that information from the source code and give a natural language response.  Q1 and Q4 correspond to the return type and signature of the method.  

The model performed worse for Q2 and Q3, in which the model identifies the method parameters and definition.  These information needs may be more difficult to learn because they vary more in size and vocabulary.  The return type (Q1) can always be found in the same place at the start of the method signature, it is generally one word long, and the vocabulary is limited to type names.  The parameter list can also always be found in the same place, but it varies in length and includes identifier names that may be specific to that method.  

The model performs the least well on Q5, especially in terms of relevance and completeness.  This result may be expected, however, since the model is expected to provide a short description of the method's behavior.  We give the model a short description in the context, and the model does learn to use this in its response.  But the size, quality,  and vocabulary of the description vary considerably, as the source is JavaDocs, making the model prone to using incorrect words.

The responses to Q6 are a special case since they are always either ``yes'' or ``no,'' making it easy for the model to score well in terms of e.g. relevance.  Still, the model is sometimes wrong, which is reflected in column Q6 of Figure~\ref{fig:rq2_boxplots}(b).

\begin{figure}[h!]
	\vspace{0.1cm}
	\setcounter{figure}{0}
	\renewcommand{\figurename}{Example}
	\vspace{-0.2cm} 
	{\small	
		\begin{tabular}{lm{6cm}}
			\emph{user query}   & Give me the return type \\
			\emph{model output} & the return type for this method is vertex \\
			\emph{context}      & \begin{footnotesize}
				\begin{verbatim}
				<st> returns the next vertex of a  
				polygon nl public vertex nextvertex ( 
				vertex v ) { int ind = vertices . 
				indexof ( v ) ; return ( vertex ) ( ind
				== -1 ? null : vertices . get ( ( ind 
				+ 1 ) % vertices . size () ) ) ; } <et>\end{verbatim}\end{footnotesize}
		\end{tabular}

		\vspace{-0.2cm}
		
		\begin{tabular}{lp{0.8cm}p{.1mm}p{0.1mm}p{0.1mm}p{0.1mm}p{0.1mm}p{0.1mm}p{0.1mm}p{0.1mm}p{0.1mm}p{0.1mm}p{0.1mm}p{0.1mm}p{0.1mm}p{0.1mm}p{0.1mm}p{0.1mm}p{0.1mm}}
			\multicolumn{2}{l}{\textless st\textgreater}	& 1  & \multicolumn{16}{l}{\multirow{12}{*}{\includegraphics[width=6cm]{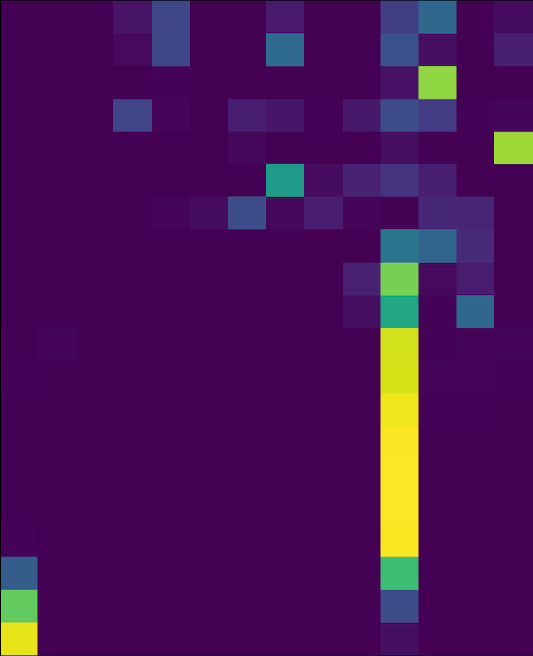}}}          \\[0.7pt]
			\multicolumn{2}{l}{the}	& 2  & \multicolumn{16}{l}{}                            \\[0.7pt]
			\multicolumn{2}{l}{return}	& 3  & \multicolumn{16}{l}{}                            \\[0.7pt]
			\multicolumn{2}{l}{type}	& 4  & \multicolumn{16}{l}{}                            \\[0.7pt]
			\multicolumn{2}{l}{for}	& 5  & \multicolumn{16}{l}{}                            \\[0.7pt]
			\multicolumn{2}{l}{this}	& 6  & \multicolumn{16}{l}{}                            \\[0.7pt]
			\multicolumn{2}{l}{method}& 7  & \multicolumn{16}{l}{}                            \\[0.7pt]
			\multicolumn{2}{l}{is}& 8  & \multicolumn{16}{l}{}                            \\[0.7pt]
			\multirow{9}{*}[17pt]{\Bigg\downarrow} & \multirow{10}{*}[23pt]{\makecell{\hspace{-0.3cm}\emph{predicting} \\ \hspace{-0.3cm}\emph{next word}}} 	  & 9  & \multicolumn{16}{l}{}                            \\[0.7pt]
			&	& 10  & \multicolumn{12}{l}{}                            \\[0.7pt]
			&	& 11 & \multicolumn{12}{l}{}                            \\[0.7pt]
			&	& 12 & \multicolumn{12}{l}{}                            \\[0.7pt]
			&	& 13 & \multicolumn{12}{l}{}                            \\[0.7pt]
			&	& 14 & \multicolumn{12}{l}{}                            \\[0.7pt]
			&	& 15 & \multicolumn{12}{l}{}                            \\[0.7pt]
			&	& 16 & \multicolumn{12}{l}{}                            \\[0.7pt]
			&	& 17 & \multicolumn{12}{l}{}                            \\[0.7pt]
			&	& 18 & \multicolumn{12}{l}{}                            \\[0.7pt]
			&	& 19 & \multicolumn{12}{l}{}                            \\[0.7pt]
			&	& 20 & \multicolumn{12}{l}{}                            \\[0.8pt]
			&	&    &1&2&3&4&5&6&7&8&9&10&11&12&13&14&&
		\end{tabular}
	}
	\vspace{-0.1cm}
	\caption{Participant asks for the return type of the method.  A heatmap of the attention network shows how the model attends to the correct position of "vertex" in the context (position 11) to predict the last word in the output.}
	\label{fig:ex1}
\vspace{-0.3cm}
\end{figure}

\subsection{RQ$_3$: Effects of Context Features}

We provide evidence of the effects of the features in the context via an example of the model's behavior.  ``Explainable AI'' is a controversial topic, with much agreement that it is necessary but little consensus on the best strategies -- neural networks in particular have a reputation for producing results that are hard to explain~\cite{ras2018explanation, samek2017explainable}.  However, one source of evidence is the attention network.  The attention mechanisms in most encoder-decoder models are responsible for connecting pieces of the decoder inputs to pieces of the encoder inputs.  Frequently attention provides clues as to why the model makes a particular decision.  E.g., in NMT the German word ``hund'' will receive high attention to the English word ``dog'', while in computer vision the word ``dog'' may receive high attention to the area in an image where a dog appears.

In our approach, the attention mechanism connects output words to words in the input context sequence.  Consider Example~\ref{fig:ex1}.  The user study participant writes a query requesting the return type.  The heatmap shows the state of the attention network just prior to predicting the word ``vertex'' (layer {\small \texttt{code\_attn}} from Section~\ref{sec:impl},  recall from Section~\ref{sec:training} that the model predicts output one word at a time).  This example is typical of almost all queries about the return type: the model has learned where to find the return type in code.  It is not always in position 11, but the model has learned to look for the signature, and where to look in the signature for the return type.  Note that the model is not attending to earlier uses of the word vertex in the method description, since that wording may change.  Likewise, it is not attending to the word vertex in after the actual return in the code, since that is a variable name which may not be the actual return type.

We include several more examples in our online appendix cited below. The behavior is quite consistent: for queries about e.g. parameters, the model attends to the parameters area of the signature, and outputs the relevant information.

\vspace{-0.1cm}
\section{Conclusion}
\label{sec:repro}

We presented a QA system for programmer questions about subroutines.  We designed a neural model based on the encoder-decoder structure that can extract information about Java methods directly from raw source code.  Our system distinguishes between and answers questions for six different information needs derived from recent related work on dialogue systems for programmers.  In an experiment with 20 professional programmers, we show that our approach is able to reliably answer these question types.

Throughout our paper, we note that this QA system is not intended for use on its own.  Instead, it would serve as a component of a hypothetical much larger interactive dialogue system.  Virtual agents are anticipated for many tasks including as assistants for software engineering.  However, it is unreasonable to expect to create such a system in one step -- research into subsystems and supporting components is required first.  This paper fills that role towards virtual agents for SE tasks.  Important next steps include both designing other subsystems and expanding the number of question types that this QA system is able to handle.

To promote continued research, we release all our data, approach source code, and a working interactive demonstration via our online appendix:

\textbf{https://github.com/paqs2020/paqs2020}

\vspace{-0.1cm}

\section* {Acknowledgments}
This work is supported in part by the NSF CCF-1452959 and CCF-1717607 grants. Any opinions, findings, and conclusions expressed herein are the authors’ and do not necessarily reflect those of the sponsors.

\bibliographystyle{IEEEtran}
\bibliography{main}

\begin{thebibliography}{10}
\providecommand{\url}[1]{#1}
\csname url@samestyle\endcsname
\providecommand{\newblock}{\relax}
\providecommand{\bibinfo}[2]{#2}
\providecommand{\BIBentrySTDinterwordspacing}{\spaceskip=0pt\relax}
\providecommand{\BIBentryALTinterwordstretchfactor}{4}
\providecommand{\BIBentryALTinterwordspacing}{\spaceskip=\fontdimen2\font plus
\BIBentryALTinterwordstretchfactor\fontdimen3\font minus
  \fontdimen4\font\relax}
\providecommand{\BIBforeignlanguage}[2]{{%
\expandafter\ifx\csname l@#1\endcsname\relax
\typeout{** WARNING: IEEEtran.bst: No hyphenation pattern has been}%
\typeout{** loaded for the language `#1'. Using the pattern for}%
\typeout{** the default language instead.}%
\else
\language=\csname l@#1\endcsname
\fi
#2}}
\providecommand{\BIBdecl}{\relax}
\BIBdecl

\bibitem{yin2016neural}
J.~Yin, X.~Jiang, Z.~Lu, L.~Shang, H.~Li, and X.~Li, ``Neural generative
  question answering,'' in \emph{Proceedings of the Twenty-Fifth International
  Joint Conference on Artificial Intelligence}.\hskip 1em plus 0.5em minus
  0.4em\relax AAAI Press, 2016, pp. 2972--2978.

\bibitem{malinowski2015ask}
M.~Malinowski, M.~Rohrbach, and M.~Fritz, ``Ask your neurons: A neural-based
  approach to answering questions about images,'' in \emph{Proceedings of the
  IEEE international conference on computer vision}, 2015, pp. 1--9.

\bibitem{weston2015towards}
J.~Weston, A.~Bordes, S.~Chopra, A.~M. Rush, B.~van Merri{\"e}nboer, A.~Joulin,
  and T.~Mikolov, ``Towards ai-complete question answering: A set of
  prerequisite toy tasks,'' \emph{arXiv preprint arXiv:1502.05698}, 2015.

\bibitem{liu2017survey}
W.~Liu, Z.~Wang, X.~Liu, N.~Zeng, Y.~Liu, and F.~E. Alsaadi, ``A survey of deep
  neural network architectures and their applications,'' \emph{Neurocomputing},
  vol. 234, pp. 11--26, 2017.

\bibitem{chen2017survey}
H.~Chen, X.~Liu, D.~Yin, and J.~Tang, ``A survey on dialogue systems: Recent
  advances and new frontiers,'' \emph{Acm Sigkdd Explorations Newsletter},
  vol.~19, no.~2, pp. 25--35, 2017.

\bibitem{gao2019neural}
J.~Gao, M.~Galley, L.~Li \emph{et~al.}, ``Neural approaches to conversational
  ai,'' \emph{Foundations and Trends{\textregistered} in Information
  Retrieval}, vol.~13, no. 2-3, pp. 127--298, 2019.

\bibitem{eberhart2020apizateaser}
Z.~Eberhart, A.~Bansal, and C.~McMillan, ``The apiza corpus: Api usage
  dialogues with a simulated virtual assistant,'' 2020.

\bibitem{samek2017explainable}
W.~Samek, T.~Wiegand, and K.-R. M{\"u}ller, ``Explainable artificial
  intelligence: Understanding, visualizing and interpreting deep learning
  models,'' \emph{arXiv preprint arXiv:1708.08296}, 2017.

\bibitem{ras2018explanation}
G.~Ras, M.~van Gerven, and P.~Haselager, ``Explanation methods in deep
  learning: Users, values, concerns and challenges,'' in \emph{Explainable and
  Interpretable Models in Computer Vision and Machine Learning}.\hskip 1em plus
  0.5em minus 0.4em\relax Springer, 2018, pp. 19--36.

\bibitem{robillard2017demand}
M.~P. Robillard, A.~Marcus, C.~Treude, G.~Bavota, O.~Chaparro, N.~Ernst, M.~A.
  Gerosa, M.~Godfrey, M.~Lanza, M.~Linares-V{\'a}squez \emph{et~al.},
  ``On-demand developer documentation,'' in \emph{2017 IEEE International
  Conference on Software Maintenance and Evolution (ICSME)}.\hskip 1em plus
  0.5em minus 0.4em\relax IEEE, 2017, pp. 479--483.

\bibitem{ward2016challenges}
N.~G. Ward and D.~DeVault, ``Challenges in building highly-interactive dialog
  systems,'' \emph{AI Magazine}, vol.~37, no.~4, pp. 7--18, 2016.

\bibitem{johnson2019no}
M.~Johnson and A.~Vera, ``No ai is an island: The case for teaming
  intelligence,'' \emph{AI Magazine}, vol.~40, no.~1, pp. 16--28, 2019.

\bibitem{rieser2011reinforcement}
V.~Rieser and O.~Lemon, \emph{Reinforcement learning for adaptive dialogue
  systems: a data-driven methodology for dialogue management and natural
  language generation}.\hskip 1em plus 0.5em minus 0.4em\relax Springer Science
  \& Business Media, 2011.

\bibitem{chen2019bidirectional}
Y.~Chen, L.~Wu, and M.~J. Zaki, ``Bidirectional attentive memory networks for
  question answering over knowledge bases,'' in \emph{Proceedings of the 2019
  Conference of the North American Chapter of the Association for Computational
  Linguistics: Human Language Technologies, Volume 1 (Long and Short Papers},
  2019, pp. 2913--2923.

\bibitem{li2015gated}
Y.~Li, D.~Tarlow, M.~Brockschmidt, and R.~Zemel, ``Gated graph sequence neural
  networks,'' \emph{arXiv preprint arXiv:1511.05493}, 2015.

\bibitem{graves2016hybrid}
A.~Graves, G.~Wayne, M.~Reynolds, T.~Harley, I.~Danihelka,
  A.~Grabska-Barwi{\'n}ska, S.~G. Colmenarejo, E.~Grefenstette, T.~Ramalho,
  J.~Agapiou \emph{et~al.}, ``Hybrid computing using a neural network with
  dynamic external memory,'' \emph{Nature}, vol. 538, no. 7626, p. 471, 2016.

\bibitem{xiong2016dynamic}
C.~Xiong, S.~Merity, and R.~Socher, ``Dynamic memory networks for visual and
  textual question answering,'' in \emph{International conference on machine
  learning}, 2016, pp. 2397--2406.

\bibitem{lemon2011learning}
O.~Lemon, ``Learning what to say and how to say it: Joint optimisation of
  spoken dialogue management and natural language generation,'' \emph{Computer
  Speech \& Language}, vol.~25, no.~2, pp. 210--221, 2011.

\bibitem{wood2018detecting}
A.~Wood, P.~Rodeghero, A.~Armaly, and C.~McMillan, ``Detecting speech act types
  in developer question/answer conversations during bug repair,'' in
  \emph{Proceedings of the 2018 26th ACM Joint Meeting on European Software
  Engineering Conference and Symposium on the Foundations of Software
  Engineering}.\hskip 1em plus 0.5em minus 0.4em\relax ACM, 2018, pp. 491--502.

\bibitem{kumar2017dialogue}
H.~Kumar, A.~Agarwal, R.~Dasgupta, S.~Joshi, and A.~Kumar, ``Dialogue act
  sequence labeling using hierarchical encoder with crf,'' \emph{AAAI}, 2018.

\bibitem{chen2018dialogue}
Z.~Chen, R.~Yang, Z.~Zhao, D.~Cai, and X.~He, ``Dialogue act recognition via
  crf-attentive structured network,'' in \emph{The 41st International ACM SIGIR
  Conference on Research \& Development in Information Retrieval}.\hskip 1em
  plus 0.5em minus 0.4em\relax ACM, 2018, pp. 225--234.

\bibitem{blunsom2013recurrent}
P.~Blunsom, N.~Kalchbrenner, and N.~Kalchbrenner, ``Recurrent convolutional
  neural networks for discourse compositionality,'' in \emph{Proceedings of the
  2013 Workshop on Continuous Vector Space Models and their
  Compositionality}.\hskip 1em plus 0.5em minus 0.4em\relax Proceedings of the
  2013 Workshop on Continuous Vector Space Models and their Compositionality,
  2013.

\bibitem{burger2000verbmobil}
S.~Burger, K.~Weilhammer, F.~Schiel, and H.~G. Tillmann, ``Verbmobil data
  collection and annotation,'' in \emph{Verbmobil: Foundations of
  speech-to-speech translation}.\hskip 1em plus 0.5em minus 0.4em\relax
  Springer, 2000, pp. 537--549.

\bibitem{he2018decoupling}
H.~He, D.~Chen, A.~Balakrishnan, and P.~Liang, ``Decoupling strategy and
  generation in negotiation dialogues,'' in \emph{Proceedings of the 2018
  Conference on Empirical Methods in Natural Language Processing}, 2018, pp.
  2333--2343.

\bibitem{Reiter:2000:BNL:331955}
E.~Reiter and R.~Dale, \emph{Building natural language generation
  systems}.\hskip 1em plus 0.5em minus 0.4em\relax New York, NY, USA: Cambridge
  University Press, 2000.

\bibitem{deng2018deep}
L.~Deng and Y.~Liu, \emph{Deep Learning in Natural Language Processing}.\hskip
  1em plus 0.5em minus 0.4em\relax Springer, 2018.

\bibitem{maalej2013patterns}
W.~Maalej and M.~P. Robillard, ``Patterns of knowledge in api reference
  documentation,'' \emph{IEEE Transactions on Software Engineering}, vol.~39,
  no.~9, pp. 1264--1282, 2013.

\bibitem{meng2018application}
M.~Meng, S.~Steinhardt, and A.~Schubert, ``Application programming interface
  documentation: what do software developers want?'' \emph{Journal of Technical
  Writing and Communication}, vol.~48, no.~3, pp. 295--330, 2018.

\bibitem{maalej2014comprehension}
W.~Maalej, R.~Tiarks, T.~Roehm, and R.~Koschke, ``On the comprehension of
  program comprehension,'' \emph{ACM Transactions on Software Engineering and
  Methodology (TOSEM)}, vol.~23, no.~4, p.~31, 2014.

\bibitem{monperrus2012should}
M.~Monperrus, M.~Eichberg, E.~Tekes, and M.~Mezini, ``What should developers be
  aware of? an empirical study on the directives of api documentation,''
  \emph{Empirical Software Engineering}, vol.~17, no.~6, pp. 703--737, 2012.

\bibitem{aghajani2018large}
E.~Aghajani, C.~Nagy, G.~Bavota, and M.~Lanza, ``A large-scale empirical study
  on linguistic antipatterns affecting apis,'' in \emph{2018 IEEE International
  Conference on Software Maintenance and Evolution (ICSME)}.\hskip 1em plus
  0.5em minus 0.4em\relax IEEE, 2018, pp. 25--35.

\bibitem{head2018not}
A.~Head, C.~Sadowski, E.~Murphy-Hill, and A.~Knight, ``When not to comment:
  questions and tradeoffs with api documentation for c++ projects,'' in
  \emph{Proceedings of the 40th International Conference on Software
  Engineering}.\hskip 1em plus 0.5em minus 0.4em\relax ACM, 2018, pp. 643--653.

\bibitem{robillard2011field}
M.~P. Robillard and R.~Deline, ``A field study of api learning obstacles,''
  \emph{Empirical Software Engineering}, vol.~16, no.~6, pp. 703--732, 2011.

\bibitem{arnaoudova2015use}
V.~Arnaoudova, S.~Haiduc, A.~Marcus, and G.~Antoniol, ``The use of text
  retrieval and natural language processing in software engineering,'' in
  \emph{Proceedings of the 37th International Conference on Software
  Engineering-Volume 2}.\hskip 1em plus 0.5em minus 0.4em\relax IEEE Press,
  2015, pp. 949--950.

\bibitem{tian2017apibot}
Y.~Tian, F.~Thung, A.~Sharma, and D.~Lo, ``Apibot: Question answering bot for
  api documentation,'' in \emph{2017 32nd IEEE/ACM International Conference on
  Automated Software Engineering (ASE)}.\hskip 1em plus 0.5em minus 0.4em\relax
  IEEE, 2017, pp. 153--158.

\bibitem{ko2004designing}
A.~J. Ko and B.~A. Myers, ``Designing the whyline: a debugging interface for
  asking questions about program behavior,'' in \emph{Proceedings of the SIGCHI
  conference on Human factors in computing systems}.\hskip 1em plus 0.5em minus
  0.4em\relax ACM, 2004, pp. 151--158.

\bibitem{pruski2015tiqi}
P.~Pruski, S.~Lohar, W.~Goss, A.~Rasin, and J.~Cleland-Huang, ``Tiqi: answering
  unstructured natural language trace queries,'' \emph{Requirements
  Engineering}, vol.~20, no.~3, pp. 215--232, 2015.

\bibitem{bradley2018context}
N.~C. Bradley, T.~Fritz, and R.~Holmes, ``Context-aware conversational
  developer assistants,'' in \emph{Proceedings of the 40th International
  Conference on Software Engineering}.\hskip 1em plus 0.5em minus 0.4em\relax
  ACM, 2018, pp. 993--1003.

\bibitem{openAPI}
H.~Ed-Douibi, G.~Daniel, and J.~Cabot, ``Openapi bot: A chatbot to help you
  understand rest apis,'' in \emph{Web Engineering}, M.~Bielikova, T.~Mikkonen,
  and C.~Pautasso, Eds.\hskip 1em plus 0.5em minus 0.4em\relax Cham: Springer
  International Publishing, 2020, pp. 538--542.

\bibitem{beringer2002promise}
N.~Beringer, U.~Kartal, K.~Louka, F.~Schiel, U.~T{\"u}rk \emph{et~al.},
  ``Promise: A procedure for multimodal interactive system evaluation,'' in
  \emph{Proceedings of the Workshop’Multimodal Resources and Multimodal
  Systems Evaluation}.\hskip 1em plus 0.5em minus 0.4em\relax Citeseer, 2002,
  pp. 90--95.

\bibitem{walker1997evaluating}
M.~A. Walker, D.~J. Litman, C.~A. Kamm, and A.~Abella, ``Evaluating interactive
  dialogue systems: Extending component evaluation to integrated system
  evaluation,'' in \emph{Interactive Spoken Dialog Systems: Bringing Speech and
  NLP Together in Real Applications}, 1997.

\bibitem{chen2019graphflow}
Y.~Chen, L.~Wu, and M.~J. Zaki, ``Graphflow: Exploiting conversation flow with
  graph neural networks for conversational machine comprehension,'' \emph{arXiv
  preprint arXiv:1908.00059}, 2019.

\bibitem{lin2019task}
Z.~Lin, X.~Huang, F.~Ji, H.~Chen, and Y.~Zhang, ``Task-oriented conversation
  generation using heterogeneous memory networks,'' in \emph{Proceedings of the
  2019 Conference on Empirical Methods in Natural Language Processing}, 2019.

\bibitem{bahdanau2014neural}
D.~Bahdanau, K.~Cho, and Y.~Bengio, ``Neural machine translation by jointly
  learning to align and translate,'' \emph{arXiv preprint arXiv:1409.0473},
  2014.

\bibitem{young2018recent}
T.~Young, D.~Hazarika, S.~Poria, and E.~Cambria, ``Recent trends in deep
  learning based natural language processing,'' \emph{ieee Computational
  intelligenCe magazine}, vol.~13, no.~3, pp. 55--75, 2018.

\bibitem{shrestha2019review}
A.~Shrestha and A.~Mahmood, ``Review of deep learning algorithms and
  architectures,'' \emph{IEEE Access}, vol.~7, pp. 53\,040--53\,065, 2019.

\bibitem{pouyanfar2018survey}
S.~Pouyanfar, S.~Sadiq, Y.~Yan, H.~Tian, Y.~Tao, M.~P. Reyes, M.-L. Shyu, S.-C.
  Chen, and S.~Iyengar, ``A survey on deep learning: Algorithms, techniques,
  and applications,'' \emph{ACM Computing Surveys (CSUR)}, vol.~51, no.~5,
  p.~92, 2018.

\bibitem{yu2020crossing}
W.~Yu, L.~Wu, Q.~Zeng, Y.~Deng, S.~Tao, and M.~Jiang, ``Crossing variational
  autoencoders for answer retrieval,'' in \emph{Proceedings of the 58th Annual
  Meeting of the Association for Computational Linguistics}, 2020, pp.
  5635--5641.

\bibitem{hayati2018retrieval}
S.~A. Hayati, R.~Olivier, P.~Avvaru, P.~Yin, A.~Tomasic, and G.~Neubig,
  ``Retrieval-based neural code generation,'' in \emph{Proceedings of the 2018
  Conference on Empirical Methods in Natural Language Processing}, 2018, pp.
  925--930.

\bibitem{leclair2019neural}
A.~LeClair, S.~Jiang, and C.~McMillan, ``A neural model for generating natural
  language summaries of program subroutines,'' in \emph{Proceedings of the 41st
  International Conference on Software Engineering}.\hskip 1em plus 0.5em minus
  0.4em\relax IEEE Press, 2019, pp. 795--806.

\bibitem{leclair2020improved}
A.~LeClair, S.~Haque, L.~Wu, and C.~McMillan, ``Improved code summarization via
  a graph neural network,'' \emph{arXiv preprint arXiv:2004.02843}, 2020.

\bibitem{chen2019sequencer}
Z.~Chen, S.~J. Kommrusch, M.~Tufano, L.-N. Pouchet, D.~Poshyvanyk, and
  M.~Monperrus, ``Sequencer: Sequence-to-sequence learning for end-to-end
  program repair,'' \emph{IEEE Transactions on Software Engineering}, 2019.

\bibitem{Linstead2009}
E.~Linstead, S.~Bajracharya, T.~Ngo, P.~Rigor, C.~Lopes, and P.~Baldi,
  ``Sourcerer: mining and searching internet-scale software repositories,''
  \emph{Data Mining and Knowledge Discovery}, vol.~18, pp. 300--336, 2009.

\bibitem{leclair2019recommendations}
A.~LeClair and C.~McMillan, ``Recommendations for datasets for source code
  summarization,'' in \emph{Proceedings of the 2019 Conference of the North
  American Chapter of the Association for Computational Linguistics: Human
  Language Technologies, Volume 1 (Long and Short Papers)}, 2019, pp.
  3931--3937.

\bibitem{craggs2005evaluating}
R.~Craggs and M.~M. Wood, ``Evaluating discourse and dialogue coding schemes,''
  \emph{Computational Linguistics}, vol.~31, no.~3, pp. 289--296, 2005.

\bibitem{wiese2017neural}
G.~Wiese, D.~Weissenborn, and M.~Neves, ``Neural domain adaptation for
  biomedical question answering,'' \emph{arXiv preprint arXiv:1706.03610},
  2017.

\bibitem{castelli2019techqa}
V.~Castelli, R.~Chakravarti, S.~Dana, A.~Ferritto, R.~Florian, M.~Franz,
  D.~Garg, D.~Khandelwal, S.~McCarley, M.~McCawley \emph{et~al.}, ``The techqa
  dataset,'' \emph{arXiv preprint arXiv:1911.02984}, 2019.

\bibitem{zhao2018finding}
H.~J. Zhao and J.~Liu, ``Finding answers from the word of god: Domain
  adaptation for neural networks in biblical question answering,'' in
  \emph{2018 International Joint Conference on Neural Networks (IJCNN)}.\hskip
  1em plus 0.5em minus 0.4em\relax IEEE, 2018, pp. 1--8.

\bibitem{code2seq}
\BIBentryALTinterwordspacing
U.~Alon, O.~Levy, and E.~Yahav, ``code2seq: Generating sequences from
  structured representations of code,'' \emph{CoRR}, vol. abs/1808.01400, 2018.
  [Online]. Available: \url{http://arxiv.org/abs/1808.01400}
\BIBentrySTDinterwordspacing

\bibitem{code2vec}
\BIBentryALTinterwordspacing
U.~Alon, M.~Zilberstein, O.~Levy, and E.~Yahav, ``code2vec: Learning
  distributed representations of code,'' \emph{CoRR}, vol. abs/1803.09473,
  2018. [Online]. Available: \url{http://arxiv.org/abs/1803.09473}
\BIBentrySTDinterwordspacing

\bibitem{multimodal}
\BIBentryALTinterwordspacing
Y.~Wan, J.~Shu, Y.~Sui, G.~Xu, Z.~Zhao, J.~Wu, and P.~S. Yu, ``Multi-modal
  attention network learning for semantic source code retrieval,'' in
  \emph{Proceedings of the 34th IEEE/ACM International Conference on Automated
  Software Engineering}, ser. ASE '19.\hskip 1em plus 0.5em minus 0.4em\relax
  IEEE Press, 2019, p. 13–25. [Online]. Available:
  \url{https://doi.org/10.1109/ASE.2019.00012}
\BIBentrySTDinterwordspacing

\bibitem{lamb2016professor}
A.~M. Lamb, A.~G. A.~P. Goyal, Y.~Zhang, S.~Zhang, A.~C. Courville, and
  Y.~Bengio, ``Professor forcing: A new algorithm for training recurrent
  networks,'' in \emph{Advances In Neural Information Processing Systems},
  2016, pp. 4601--4609.

\bibitem{logar1993comparison}
A.~M. Logar, E.~M. Corwin, and W.~J. Oldham, ``A comparison of recurrent neural
  network learning algorithms,'' in \emph{IEEE International Conference on
  Neural Networks}.\hskip 1em plus 0.5em minus 0.4em\relax IEEE, 1993, pp.
  1129--1134.

\bibitem{doya2003recurrent}
K.~Doya, ``Recurrent networks: learning algorithms,'' \emph{The Handbook of
  Brain Theory and Neural Networks,}, pp. 955--960, 2003.

\bibitem{sridhara2011automatically}
G.~Sridhara, L.~Pollock, and K.~Vijay-Shanker, ``Automatically detecting and
  describing high level actions within methods,'' in \emph{Proceedings of the
  33rd International Conference on Software Engineering}.\hskip 1em plus 0.5em
  minus 0.4em\relax ACM, 2011, pp. 101--110.

\bibitem{mcburney2016automatic}
P.~W. McBurney and C.~McMillan, ``Automatic source code summarization of
  context for java methods,'' \emph{IEEE Transactions on Software Engineering},
  vol.~42, no.~2, pp. 103--119, 2016.

\bibitem{sridhara2010towards}
G.~Sridhara, E.~Hill, D.~Muppaneni, L.~Pollock, and K.~Vijay-Shanker, ``Towards
  automatically generating summary comments for java methods,'' in
  \emph{Proceedings of the IEEE/ACM international conference on Automated
  software engineering}.\hskip 1em plus 0.5em minus 0.4em\relax ACM, 2010, pp.
  43--52.

\bibitem{yin2017case}
R.~K. Yin, \emph{Case study research and applications: Design and
  methods}.\hskip 1em plus 0.5em minus 0.4em\relax Sage publications, 2017.

\end{thebibliography}

\end{document}